# LED-it-GO
## Leaking (a lot of) Data from Air-Gapped Computers via the (small) Hard Drive LED

Mordechai Guri, Boris Zadov, Eran Atias, Yuval Elovici

Ben-Gurion University of the Negev

Cyber Security Research Center

gurim@post.bgu.ac.il; borisza@gmail.com; elovici@bgu.ac.il

Video: https://www.youtube.com/watch?v=4vIu8ld68fc

*Abstract*

In this paper we present a method which allows attackers to covertly leak data from isolated, air-gapped computers. Our method utilizes the hard disk drive (HDD) activity LED which exists in most of today's desktop PCs, laptops and servers. We show that a malware can indirectly control the HDD LED, turning it on and off rapidly (up to 5800 blinks per second) – a rate that exceeds the visual perception capabilities of humans. Sensitive information can be encoded and leaked over the LED signals, which can then be received remotely by different kinds of cameras and light sensors. Compared to other LED methods, our method is unique, because it is also *covert* - the HDD activity LED routinely flickers frequently, and therefore the user may not be suspicious to changes in its activity. We discuss attack scenarios and present the necessary technical background regarding the HDD LED and its hardware control. We also present various data modulation methods and describe the implementation of a user-level malware, that doesn't require a kernel component. During the evaluation, we examine the physical characteristics of different colored HDD LEDs (red, blue, and white) and tested different types of receivers: remote cameras, 'extreme' cameras, security cameras, smartphone cameras, drone cameras, and optical sensors. Finally, we discuss hardware and software countermeasures for such a threat. Our experiment shows that sensitive data can be successfully leaked from air-gapped computers via the HDD LED at a maximum bit rate of 4000 bit/s (bits per second), depending on the type of receiver and its distance from the transmitter. Notably, this speed is 10 times faster than the existing optical covert channels for air-gapped computers. These rates allow fast exfiltration of encryption keys, keystroke logging, and text and binary files.

## I. INTRODUCTION

In the modern cyber era, attackers have proven that they can breach many organizations thought to be secured. They employ sophisticated social engineering methods and exploit 0-day vulnerabilities in order to infiltrate the target network, while bypassing defense measures such as intrusion detection and prevention systems (IDS/IPS), firewalls, antivirus programs, and the like. For that reason, when highly sensitive information is involved, so-called *air-gap isolation* is used. In air-gap isolation, a network is kept separate, physically and logically, from public networks such as the Internet. Air-gapped networks are commonly used in military defense systems, critical infrastructure, banks and the financial sector, and others industries [1] [2].

But despite the high degree of isolation, even air-gapped network are not immune to breaches. In recent years it has been shown that even air-gapped networks can become compromised. In order to breach such networks, attackers have used complex attack vectors, such as supply chain attacks, malicious insiders, and social engineering. While the most well-known breach cases are Stuxnet [3] and agent.btz [4], other attacks have also been reported [5] [6] [3] [7] [8].

*A. Leaking data through an air-gap*

While *infiltrating* air-gapped systems has been shown feasible in recent years, the *exfiltration* of data from systems without networking or physical access, or Internet connectivity is still considered a challenging task. Over the years, different types of air-gap covert channels (that is, covert channels aimed at leaking data from air-gapped computers) have been proposed. The electromagnetic emission from computer components is one type of covert channel that has been extensively studied. In this method, a malware controls the electromagnetic emission from computer parts, including the screen, cables, processors and other peripherals [9] [10] [11] [12] [13]. Leaking data over ultrasonic waves [14] [15] and via thermal manipulations [16] have also been studied.

A few types of optical covert channels have been presented as well. In particular, leaking data via the blinks made by keyboard LEDs, or by inserting malicious hardware with controlled LEDs into an organization. However, these methods are not considered completely covert, since they can easily be detected by people who notice the anomalous LED blinking. Generally speaking, because optical and LED methods are considered less covert, they have not received as much attention from researchers.

*B. Our contribution*

In this paper we present a method that enables malware to leak data from air-gapped computers using the HDD activity LED which is present in nearly all desktop and laptop computers today. A malware can manipulate the HDD LED and control its blinking period and speed, by using certain HDD I/O operations, such as 'read' and 'write.' We show that arbitrary data can be modulated and transmitted over the optical signals.

Compared with existing optical methods, our method is unique in five ways:

- **Covertness**. Until now, leaking data through PC LEDs has not been considered covert – given the irregular and inconstant nature of the blinking of keyboard and screen power LEDs, hence leaking data through these LEDs can be easily detected. In contrast, our method is considered covert, because unlike the keyboard and screen LEDs, HDD activity LEDs are frequently active, and manipulations of the blinking timing and speed may not draw special attention.

- **Speed**. Our measurements show that the HDD LED can be controlled and adjusted to operate at a relatively fast speed (over 4000Hz). Therefore, we were able to transmit messages at a faster speed than other LEDs methods were able to achieve. This rate allowed the exfiltration of an encryption key of 4096 bits in a matter of minutes (and even seconds), depending on the receiver.

- **Visibility**. When the HDD LED blinks for short period of time, humans may not be able to perceive its activity [17]. Moreover, at high speeds (e.g., above 400Hz), the LED flickering is invisible to humans, making the channel more covert.

- **Availability**. Our method does not require any special hardware. It works with any computer that has an HDD activity LED. This component is found on most desktop PCs, laptops, and servers today.

- **Privilege level.** Activating the HDD LED can be initiated from an ordinary user-level code, and does not required special component in the OS kernel.

The rest of the paper is organized as follows. In Section II we present related work. Section III describes the adversarial attack model. Section IV provides technical background. Data modulation and transmission are discussed in Section V. Section VI presents the evaluation and results. Countermeasures are discussed in Section VII, and we present our conclusions in Section VIII.

## II. RELATED WORK

Leaking data from air-gapped computers via covert channels has been the subject of research for the past twenty years. Air-gap covert channels can be categorized as electromagnetic, acoustic, thermal, and optical channels. In electromagnetic methods, the attacker modulates data over of the electromagnetic signals generated by various components within the computer.

### A. Electromagnetic

Back in 1998, Kuhn and Anderson [10] introduced the 'Soft Tempest' attack which involved hidden data transmission using electromagnetic emanations from a video cable. AirHopper [9], introduced in 2014, is a type of malware aimed at leaking data from air-gapped computers to a nearby mobile phone by generating FM radio signals from the video card. GSMem malware [13], presented in 2105, enables leaking data at cellular frequencies via electromagnetic emission, generated by the computer RAM bus. More recently, researchers presented USBee and Funthenna [18] which exploits the electromagnetic interference generated by the USB and GPIO buses. Matyunin et al use the magnetic field sensors of mobile devices as a covert channel [19]. Other electromagnetic and magnetic covert channels are discussed in [20].

### B. Sonic/ultrasonic

Near ultrasonic methods for air-gap covert channels are discussed in a number of academic works. Hanspach and Goetz [21] present a method for near ultrasonic covert networking using speakers and microphones in laptops. The concept of communicating over inaudible sounds has been comprehensively examined by Lee et al [22] and has also been extended for different scenarios using laptops and mobile phones [23] . Guri et al introduced Fansmitter [24] and Disfiltration [25], new methods enabling acoustic data exfiltration from computers without speakers or audio hardware. The proposed methods utilized computer fans and hard drives to generate acoustic signals.

### C. Thermal

BitWhisper [26] is a unique air-gap covert channel, allowing bidirectional covert communication between adjacent air-gapped computers, using the computers' heat emissions and the built-in thermal sensors of the computers' motherboards.

### D. Optical

In the optical domain, Loughry and Umphress [27] and Sepetnitsky and Guri [28] discuss the risks of intentional information leakage through optical signals sent from the keyboard and screen LEDs. They implemented malware that control the keyboard and screen power LEDs to transfer data to a remote camera. The main drawback of these methods is that they are less covert: since the keyboard and screen LEDs don't blink typically, users can easily detect this type of communication. Shamir et al demonstrated how to establish a covert channel with a malware over the air-gap using a blinking laser and standard all-in-one-printer [29]. However, this method is also not covert and its success relies upon user absence. More recently, Lopes and Arana [30] presented a novel approach for air-gap data exfiltration using a malicious storage device which transmits data through blinking infrared LEDs. In this way, an attacker can leak sensitive data stored on the device, such as credentials and cryptographic keys, at a speed of 15 bit/s. The computer need not be infected with a malware, but this approach does require the attacker find a way to insert the compromised hardware implanted with infrared LEDs into the organization. Brasspup [31] demonstrated how to conceal secret images in a modified LCD screen. His method required removing the polarization filter of the LCD screen which makes it less practical for real world attacks. VisiSploit [32] is another optical covert channel in which data is leaked from the LCD screen to a remote camera via a so-called 'invisible image.' With this method, a remote camera can reconstruct an invisible QR code projected on the computer screen.

Table 1 summarizes the different types of existing air-gap covert channels and presents their maximum bandwidth and effective distance.

Table 1. Different type of air-gap covert channels and distances

| Method | Examples | Max bandwidth | Effective Distance |
|---|---|---|---|
| Electromagnetic | AirHopper [9]<br>GSMem [13]<br>USBee<br>Funthenna [11] | 480 bit/s<br>1 to 1000 bit/s<br>4800 bit/s | ~5-10 meters |
| Acoustic | [15] [14] [22] [23] [33]<br>Fan noise (Fansmitter) [24]<br>Hard disk noise (DiskFiltration) [25] | <100 bit/s<br>900 bit/h<br>10K bit/h | ~15 meters |
| Thermal | BitWhisper | 1-8 bit/h | 40 cm |
| Optical (LEDs) | (unmodified) Keyboard LEDs [27]<br>Screen LEDs [28]<br>Implanted infrared LEDs [30] | 150 bit/s<br>20 bit/s<br>15 bit/s | Line of sight |
|  | Hard drive LED (this paper) | 4000 bit/s | Line of sight |

The method presented in this paper has two main advantages compared to other LED based methods: its covertness and its speed. The state of keyboard and screen power LEDs does change frequently, and it is likely that any blinking caused by tampering with this channel will be obvious to observers. On the other hand, the HDD activity LED blinks frequently (due to OS background operations), and hence the effects of communication via this channel (changes in the blinking patter) will not arouse attention. The exfiltration speed we achieved with the HDD LED is up to 4000 bit/s, much higher than existing air-gap LED methods.

### III. ADVERSARIAL ATTACK MODEL

The attack model consists of two phases: (1) infecting the target computer with a malware, and (2) receiving and decoding the signals that were leaked through the HDD LED.

**Infection**. As demonstrated in recent years, infecting a computer within a secure network can be accomplished. Attackers may employ supply chain attacks, use social engineering techniques, or launch hardware with preinstalled malware to obtain a foothold in the target machine [34] [35] [36]. The malware then gathers sensitive information from the user's computer (e.g., keystrokes, password, encryption keys, and documents). Eventually it starts transmitting the binary data through the blinking HDD LED using a selected encoding scheme.

**Reception and Decoding.** The attack model also requires a digital camera or optical sensor which has a line of sight with the compromised computer's front panel. We identify two scenarios in which this threat model is relevant: (1) the 'malicious insider' [37] (also known as the 'evil maid' [38] attack), in which a person carrying a hidden camera can obtain a line of sight with the compromised computer, and (2) a scenario involving a type of remote camera or optical sensor pointed at the compromised computer [39].

There are several types of equipment that can play the role of the receiver in this attack model.

- **Local hidden camera**. A hidden camera that has a line of sight to the front panel of the transmitting computer.
- **High resolution remote camera**. A high resolution camera (or other type of optical sensor) which is located outside the building, but positioned so it has a line of sight with front panel of the transmitting computer.
- **Drone camera**. A camera installed on a versatile drone which is flown to a location which has a line of sight with the front panel of the transmitting computer, e.g., near the window.

This type of receiver is relevant for leaking a small amount of data (e.g., leaking encryption keys).

- **Camera carried by malicious insider**. A person that stands in close proximity to the computer, and can position him/herself so as to have a line of sight with the front panel of the transmitting computer, carrying a smartphone or wearable video camera (e.g., hidden camera).
- **Compromised security camera**. A security camera positioned in a location where it has a line of sight with the front panel of the transmitting computer. A comprehensive analysis of the threats, vulnerabilities, and attacks on video surveillance, closed-circuit TV, and IP camera systems was conducted by Costin in [39].
- **Optical sensors**. An optical sensor capable of sensing the light emitted from the HDD LED. Such sensors are used extensively in VLC (visible light communication) and LED to LED communication [40]. Notably, optical sensors are capable of sampling LED signals at high rates, enabling data reception at a higher bandwidth than a typical video camera.

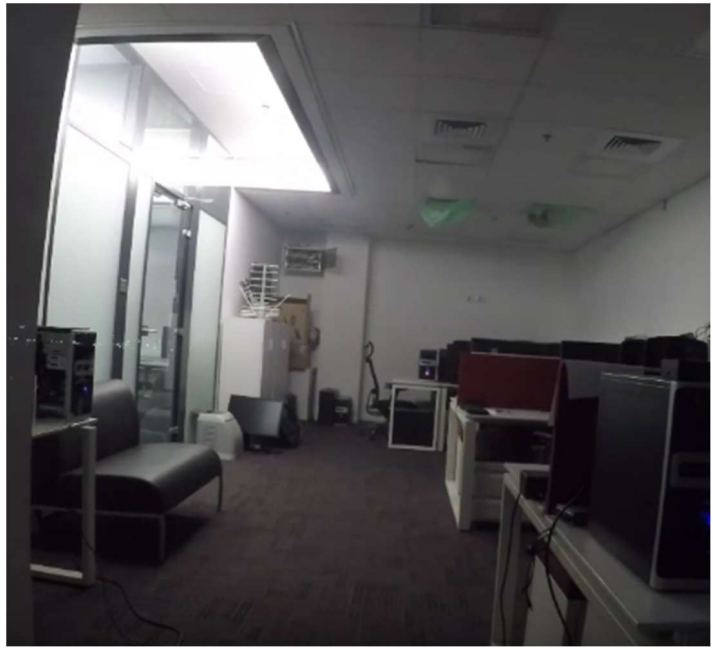

**Figure 1. Example of the attack. A leaking HDD LED (the red LED at the center of the image) as captured by a hidden video camera eight meters away from the transmitting computer.**

An example of the covert channel is provided in Figure 1 in which sensitive data is encoded in binary form and covertly transmitted over a stream of HDD LED signals. A hidden video camera films the activity in the room, including the LED signals. The attacker can then decode the signals and reconstruct the modulated data.

IV. TECHNICAL BACKGROUND

LED (light emitting diode) is a two-lead semiconductor light source that illuminates when an electrical charge passes through it. LEDs are used as activity indicators in a wide range of electronic devices. In addition, LEDs of different size and colors are also used in various applications such as advertising, home lighting, automotive industry and traffic signals. The wavelength of the emitted light (which is indicated by its color) is determined by the material used in the semiconducting element within the LED. Generally, aluminum gallium indium phosphide (AlGaInP) is used in red, orange, and yellow LEDs, and indium gallium nitride (InGaN) is used in green, blue, and white LEDs.

LED indicators are commonly used in desktop and laptop computers and their peripheral hardware components. Computer LEDs include the power and keyboard LEDs. Many motherboards have an internal onboard power LED that indicates whether there are hardware errors.

*A. Hard drive activity LED*

The HDD activity indicator is a small LED that blinks whenever the hard drive is active (being read from or written to). A schematic flow of the HDD activity LED's operation is provided in Figure 2. Technically, internal hard disk drives in desktop and laptop computers are connected to the motherboard via a SATA/IDE interface or another type of mass storage device interface (Figure 2-1). The motherboard hard drive controller sends signals via the HDD activity header pins over the standard 2-pin HDD LED extension cable (Figure 2-2). These signals cause an LED on the front of a computer to flash when the drive is active. The signals are sent to the LED whenever 'read' or 'write' operations have been issued to the HDD. Some large computer cases (e.g., server PCs) have multiple hard disk activity LEDs, to allow for separate connection of a number of drives. External hard drives and flash drives are usually equipped with an activity LED as well. In these cases, the LED is connected to the embedded controller of the flash drive.

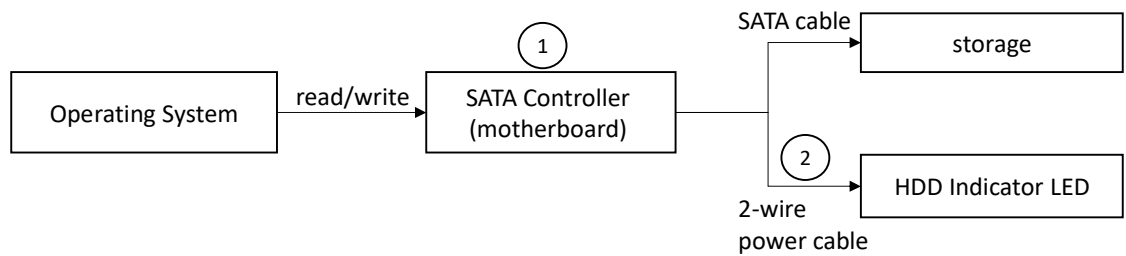

Figure 2. Schematic flow of the HDD activity LED operation.

On a desktop PC the HDD LED is located on the front panel of the computer. On a laptop the HDD LED is usually located on the front control bar (located above the keyboard) or on the front edge of the computer. The activity LED of an external HDD is usually located on the front of its case. The LED light may be any color, depending on the type of computer, but it is usually white, red, green, yellow, or blue.

*B. HDD LED circuit hardware*

The HDD activity LED is connected to the motherboard circuit using a 2-pin extension cable. To activate the component two parameters are needed: forward voltage of at least 2V, and forward current of 20-130mA, depending on type of LED. Digital output ports cannot provide an adequate amount of current needed to operate the LED, and thus an extra electronic circuit, called an LED driver, is needed to fulfill the working requirements. Figure 3 presents the two driver circuits commonly used in motherboards. The circuit in Figure 3(a) is based on an NPN transistor connected in common emitter configuration. The circuit in Figure 3(b) based on an operational amplifier. Both circuits can provide the amount of current needed to operate the LED.

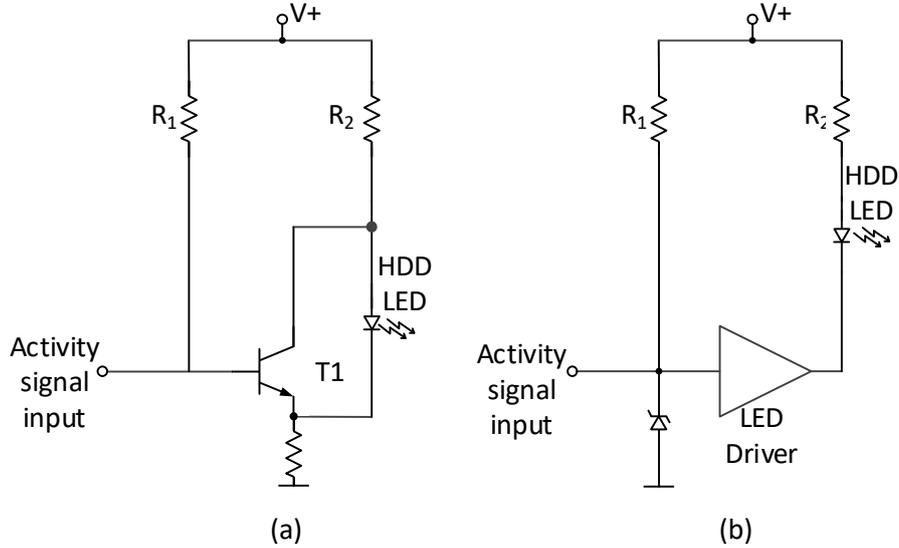

Figure 3. The two common motherboard LED driver circuits. (a) A common emitter configuration with self-biasing implementation. (b) Operational amplifier based implementation.

The circuit presented in Figure 3(a) is less expensive than the circuit in Figure 3(b), and it has more limited current amplification (150-550). The more expensive circuit in Figure 3(b) is based on an operational amplifier and has a comparator configuration with open loop amplification of $10^6$, allowing faster responsivity to the input signal; in addition it has only two voltage levels (5v and 0v). This circuit also has better immunization to the input.

## V. DATA TRANSMISSION

In this section we present the data transmission. We discuss the basic signal generation and describe different modulation methods, along with their implementation details.

*A. Signal generation*

In our method, the data carrier is the state of the HDD LED. The basic signal is generated by turning the LED on and off. Technically, the HDD LED is controlled directly by the motherboard chipset. We found that there is no reliable generic API that can be used to enable a software to request the motherboard to turn on the HDD indicator LED. In order to generate the signal, we can indirectly control the LED by performing specific HDD 'read' or 'write' operations. These operations cause the motherboard to turn on the HDD LED for a specified amount of time, depending on the size of the buffer being read or written to the storage device. Table 2 lists the OS level operations and the corresponding LED states, denoted as LED-ON and LED-OFF. As can be seen, reading or writing a buffer size $S$ causes the HDD LED to be turned on for the time period of $T_{on}$. Sleeping for time $T_{off}$ causes the HDD LED to be turned off for a period of $T_{off}$.

Table 2. Signal time

| OS operation | HDD-LED state |
|---|---|
| read/write ($S$) | LED-ON for time $T_{on}(S)$ |
| sleep ($T_{off}$) | LED-OFF for time $T_{off}$ |

*B. Data encoding*

The topic of visible light communication has been widely studied in the last decade. In particular, various modulations and encoding schemes have been proposed for LED to LED communication [40] [41] [42]. For our purposes, we present three basic encoding schemes which enable the transmission of digital data over the HDD LED: (1) on-off keying (OOK), (2) Manchester encoding, and (3) Binary Frequency Shift Keying (B-FSK).

*1) On-off keying (OOK)*

This is the simplest form of the more general amplitude-shift keying (ASK) modulation. The presence of a signal for a certain duration encodes a logical one ("1"), while its absence for the same duration encodes a logical zero ("0"). In our case, LED-ON for a duration $T_{on}$ encodes "1" and LED-OFF for duration of $T_{off}$ encodes "0". Note that in the simple case $T_{on} = T_{off}$.

*2) Manchester encoding*

In Manchester encoding each logical bit is sent using two physical bits. The sequence of physical bits "01" (LED-OFF, LED-ON) encodes a logical "0" and the sequence of physical bits "10" (LED-ON, LED-OFF) encodes a logical "1". Manchester encoding solves the LED flickering problem by sending an equal number of ones and zeroes. Manchester encoding's transfer rate is half of that of OOK, since it uses two physical bits for each logical bit. This type of encoding is considered more reliable because of the redundancy of each transmitted bit; therefore it is heavily used in communication.

*3) Binary Frequency Shift Keying (B-FSK)*

In this encoding scheme both a logical "1" and "0" are encoded by LED-ON. A logical "1" is encoded by the LED-ON state for time duration of $T_{on}(S_1)$, and a logical "0" is encoded by the LED-ON state for time duration of $T_{on}(S_0)$. Each logical bit is followed by a guard interval (LED-OFF) for a time interval of $T_{off}$.

*C. Bit framing*

We transmit the binary data in frames. The frames have two roles: (1) providing periodic synchronization signals to the receiver, and (2) providing a basic error check mechanism for each sent packet of bytes.

We used two types of bit framing: (1) fixed length framing, and (2) variable length framing.

Table 3. Fixed length framing

| 8 bits | 256 bits | 16 bits |
|---|---|---|
| **Preamble** (10101010) | Payload | CRC |

Table 4. Variable length framing

| 8 bits | 16 bits | n bits | 16 bits |
|---|---|---|---|
| **Preamble** (10101010) | Payload size (n) | Payload | CRC |

In fixed length framing (Table 3) the binary data is transmitted in small, fixed size packets. Each packet is composed of a preamble (8 bits), a payload (256 bits), and a checksum (16 bits). The **preamble** consists of a sequence of eight alternating bits ('10101010') and is used by the receiver to periodically determine the channel timing ($T_{on}$ and $T_{off}$). In addition, the preamble header allows the receiver to identify the beginning of a transmission and calibrate other parameters, such as the intensity and color of the transmitting LED. The **payload** is a chunk of 256 bits to be transmitted. We used a 16-bit **CRC** (cyclic redundancy check) for error detection.

The CRC is computed on the payload and added to the end of the frame. The receiver calculates the CRC for the received payload, and if it differs from the received CRC, an error is detected. In variable length framing (Table 4) the binary data is transmitted in packets with varying length. The preamble is followed by 16 bits which determine the payload size. The payload size may differ between packets. Finally, the 16-bit CRC of the payload is added to the end of the frame.

Note that fixed length bit framing is more suitable for cases in which a small amount of fixed sized data is about to be transmitted (e.g., encryption keys and passwords). With a larger amount of data (e.g., files), variable length framing may be better, because it can transmit the entire amount of data in fewer packets, while saving the overhead of the frame headers. However, in some circumstances larger frames are more wasteful, since a single bit error may corrupt the whole frame.

*D. Implementation*

We implemented a prototype of the transmitter for the Linux OS. We choose to use the read operation to turn on the HDD activity LED, because it leaves no traces on the file system. We executed a C program which uses the direct addressing system calls and the *fseek(), fopen(),* and *fread()* system calls [43] [44]. We also implemented a shell script version of the transmitter using the Linux dd command-line utility [45]. This is a low level utility of Linux which can perform a wide range of HDD operations (e.g., read or copy) at the file or block level. A pseudocode for on-off keying transmission is provided in algorithm 1.

**Algorithm 1** HDDLED_TransmitBit
```
1: procedure transmitBits(bits, T0, ReadSize)
2: sync(); //drop cache
3: hddDev = open(/dev/sda)
4: offset = 0
5: offsetIncrement = BLOCK_SIZE;
6: seek(hddDev, offset);
7: for(b in bits)
8:    if (b='0') then
9:       sleep (T0);
10:   if (b='1') then
11:      seek(hddDev, offset);
12:      read(hddDev, ReadSize);
13:      offset += offsetIncrement
14: end for
15: return;
```

The procedure HDDLED_TransmitBit takes three parameters: the stream of bits to transmit (*bits*), time $T_{off}$ (*T0*), and the size of the buffer for the read operation (*ReadSize*), which determines $T_{on}(S)$. Initially, the cache is cleared (line 2), and then we open the main hard drive for reading (line 3). Since the OS performs HDD reads in small size blocks (BLOCK_SIZE), we must ensure that two consecutive reads are taken on different blocks in the HDD; otherwise, the second read operation will not generate HDD access (LED activity), because the block is already in the cache. For each bit in the bit stream, if the bit is '0' we do nothing for time *T0* (line 8-9). If the bit is '1' then a read operation is performed, and we advance to the next block (line 11-13).

*E. Caching avoidance*

In order to efficiently modulate data over the LED signal, we need to precisely control the duration of the read operations. In particular, we need to be able to perform the read operation at a given time without delay. Modern OSs employ disk and file I/O caching mechanisms in their kernel or device drivers. Such caching can cause timing delays and inconsistencies in the generation of LED signals. For efficient and error-free signal generation, we must avoid any type of caching during the read operation. Before the transmissions, we turned off the disk caching using the /proc/sys/vm/drop_caches, in order to instruct the kernel to free the pagecache, dentries, and inodes. We also turned off the HDD write-back cache mechanism, using the hdparm command line tool [46]. In the shell script, we used the dd tool with 'direct' flag (use direct I/O for data), and 'sync' flag (use synchronized I/O for data).

## VI. EVALUATION

In this section we present the evaluation of the transmitter and its characteristics. Note that the concept of LED communication was the subject of a considerable amount of research in recent years. Today it is possible to transmit signals over LEDs at a rate of 500 Mbit/s [42] [47]. Other research shows that it is possible to decode LED signals from more than 30 meters away [27]. Our evaluation focuses on the characteristics of the HDD LED and its rate. In our evaluation, we adapt the approach commonly used in visible light communication (VLC), which assumes a line of sight between the light source and the camera [27] [41].

The experimental setup consists of three off the shelf standard desktop PCs, each with a different type of HDD LED (Table 5). Note that although most HDD indication LEDs are red, many vendors today are using different colors. For the evaluation we used red, blue, and white types of HDD LEDs.

Table 5. PCs used in our experimental setup with red, blue, and white HDD indicator LEDs

| # | Case type | LED | Motherboard | Hard Drive |
|---|---|---|---|---|
| **PC-1** | Infinity | Red | Gigabyte H87M-D3H | - WD Blue 1TB Desktop Hard Disk Drive - 7200 RPM SATA 6Gb/s 64MB Cache 3.5"<br>- Kingston HyperX Savage SSD, 240GB |
| **PC-2** | Gigabyte | Blue | Gigabyte H77-D3H | Seagate Desktop HDD 1TB 64MB Cache SATA 6.0Gb/s 3.5" |
| **PC-3** | Dell Optiplex 9020 | White | Intel Q87(LYNX POINT) | WD Blue 1TB Desktop Hard Disk Drive - 7200 RPM SATA 6Gb/s 64MB Cache 3.5"" |

The main PC has an Infinity case with a Gigabyte H87M-D3H motherboard with Intel® H87 chipsets. We tested two hard drives. The first is the WD Blue 1TB Desktop Hard Disk Drive - 7200 RPM SATA 6Gb/s 64MB Cache 3.5 Inch. Based on a benchmark that we conducted in our lab, this HDD has a read rate of 144.2 MB/s and an access time of 14.7ms. We also tested the 240GB Kingston HyperX Savage solid state drive (SSD). A small circular red HDD indicator LED is located on the front of the computer chassis. This LED is connected to two pins in the motherboard (HD LED pins), which supply its voltage. In terms of software we used the user-level transmitting program described in the Implementation sub-section. This program receives the channel parameters (LED-ON and LED-OFF times) and the array of bits to transmit.

### A. LED measurement setup

To evaluate the HDD LED at high speeds we designed a measurement setup based on photodiode light sensors.

**Setup**. Very simply, a photodiode is a semiconductor that converts light into electrical current. The measurement setup is shown in Figure 4. The photodiode was connected to the charge configuration amplifier AD549 [48] (with 0.11 fA current noise density) and the data acquisition system. Data was collected with the NI cDAQ measurement system with a 16-bit ADC NI9223 measurement card [49] which is capable of 200K samples per second. The system was driven by the LabVIEW dataflow visual programming language.

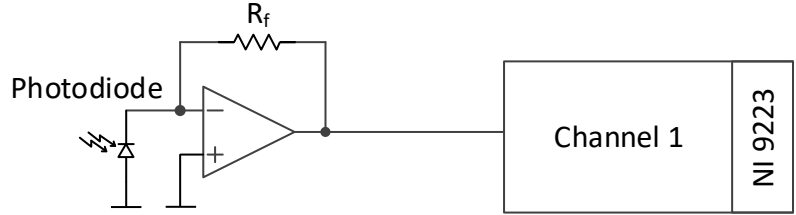

**Figure 4. The light measurement experimental setup, based on a photodiode and NI acquisition system with high speed sampling rate.**

**Sensors.** We used two type of photodiodes: (1) SIEMENS photodiode SFH-2030 (Figure 5(a)), a device which is suitable for sensing light at wavelengths of 400-1100 nm, dark current of 1nA, 55nsec of current rise and fall time, and (2) Thorlabs PDA100A Si Switchable Gain Detector [50] (Figure 5(b)), suitable for sensing light at wavelengths of 320-1100 nm, 2.4 MHz BW, 0.973 - 27 pW/Hz1/2 with a built-in amplifier.

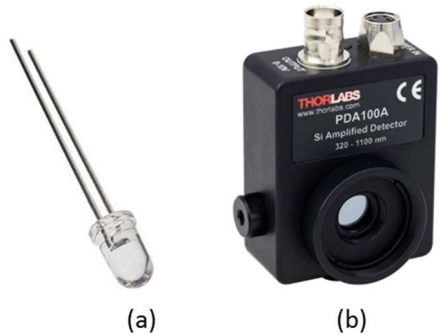

**Figure 5. (a) SIEMENS photodiode SFH-2030, and (b) Thorlabs PDA100A sensor.**

The two sensors fit into different attack models. The SIEMENS photodiode circuit is a simple, small sensor that could be secretly hidden in the room in which the transmitter computer is located. The Thorlabs PDA100A is a larger device with a built-in optimized amplifier and an option to connect an external lens used to magnify and focus the light signal. Such a larger sensor may be located remotely (e.g., outside the window), carried by human, or integrated on a flying drone.

*B. The transmitter*

As mentioned, the HDD activity LED cannot be controlled directly by software. Instead, we perform the 'read' operation from the OS, which in turn causes the HDD controller on the motherboard to turn on the HDD indication LED. We examine the correlation between $S$ and $T_{on}(S)$, denoting the number of bytes we read as $S$ and the time the LED was on as $T_{on}(S)$. This information is important, because it enables us to configure the parameters of the transmitting software. We also evaluate the channel boundaries to determine the maximum frequency on which the HDD LED can operate and the maximum bandwidth.

**Table 6. LED-ON measurements**

| Read volume ($S$) | LED-ON Time ($T_{on}(S)$) | Bit/s |
|---|---|---|
| 60000000 (60MB) | 630 ms | 1.6 bit/s |
| **15120000 (15MB)** | 250 msec | 4 bps bit/s |
| **80000000 (8MB)** | 60 msec | 16 bit/s |
| **5120000 (5MB)** | 32 msec | 30 bit/s |
| **1280000 (1.2MB)** | 5 msec | 180 bit/s |
| **800000 (800KB)** | 3.6 msec | 277 bit/s |
| **600000 (600KB)** | 3.2 msec | 312 bit/s |
| **512000 (512KB)** | 2 msec | 500 bit/s |
| **256000 (256KB)** | 1.2 msec | 833 bit/s |
| **<4KB** | 0.18 msec | 4000 bit/s |

We configured the transmitter to perform the 'read' operation with varying the number of bytes ($S$). Using the LED measurement setup described above, we measured the corresponding LED-ON time ($T_{on}(S)$). The photodiodes were positioned directly in front of the HDD LED, and located a distance from zero to several meters away. Table 6 contains the main values of $S$ and the corresponding $T_{on}(S)$.

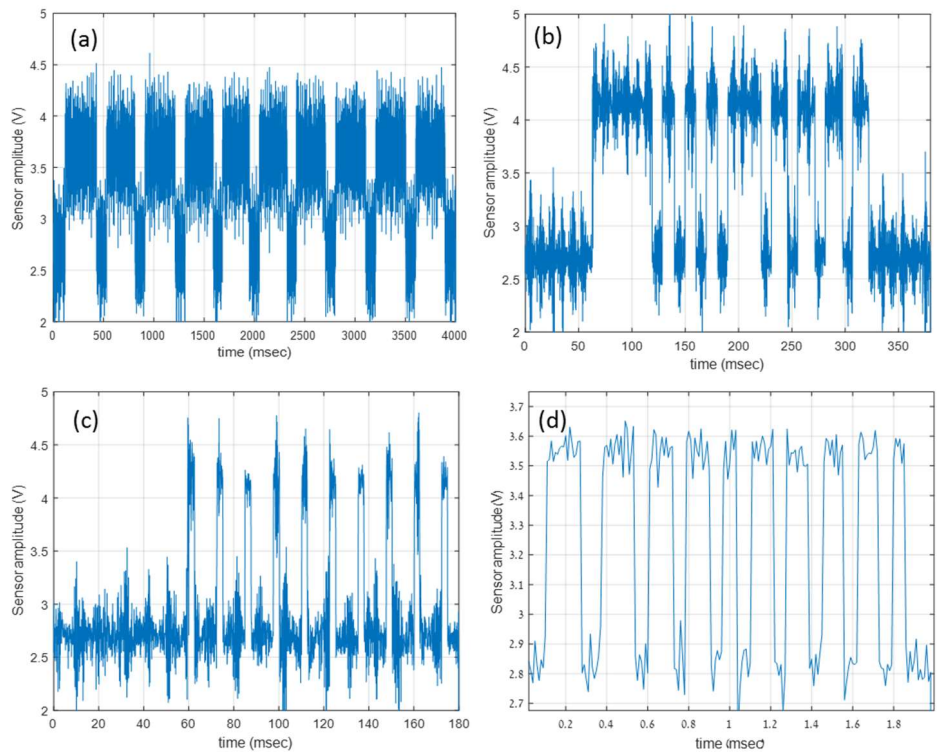

**Figure 6.** PC-1 measurements with pulse width of (a) 680 msec ($S = 60M$), (b) 32 msec ($S = 5MB$), (c) 3.6 msec ($S = 0.8MB$), and (d) 0.18 msec ($S = 4K$).

Figure 6 shows the waveforms of binary '101010…' (encoded in OOK) transmitted at a rate of 1.6 bit/s, 30 bit/s, 277 bit/s, and 4000 bit/s, respectively. In this case, the transmitter is the PC-1 (red LED), and the receiver is the SIEMENS photodiode. Note that our LED-ON measurement includes a duty cycle of 50%.

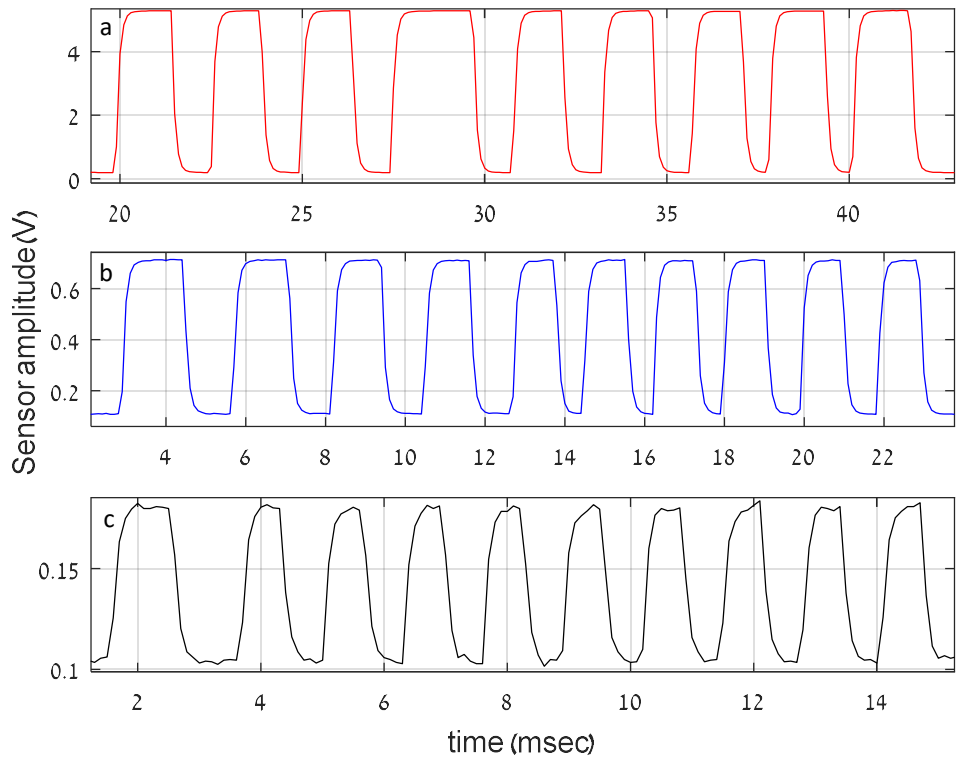

**Figure 7. Measurements for (a) PC-1 (red LED), (b) PC-2 (blue LED), and (c) PC-3 (white LED)**

Figure 7 provides measurements for transmissions made from PC-1 (red LED), PC-2 (blue LED), and PC-3 (white LED). In this experiment we use the PDA100A sensor with $S = 4K$. PC-1 has a pulse width of 0.18 msec and an amplitude 5.3V. PC-2 shows a pulse width of 0.12 msec and an amplitude of 0.71V. PC-3 shows a pulse width of 0.1 msec and an amplitude of 0.18V. The amplitude represents the amount of charge converted to voltage with the electrical current amplifier. As can be seen, the blue LEDs produce the strongest optic signals.

Table 7 shows the maximum bandwidth for the various receivers and attack models. The main factor in determining the maximum bandwidth in video cameras is the number of frames per second (FPS). For video cameras, we have identified a setting of two frames per bit as the optimal setting needed to achieve successful detection of the LED-ON timing.

**Table 7. Maximum bandwidth of different receivers**

| Tested Camera/Sensor | Model | Resolution | Max bandwidth |
|---|---|---|---|
| **Entry-level DSLR** | Nikon D7100. lens: Nikon18-140mm F3.5-5.6 ED VR | 1920x1080 (video) 1280x720 (60 fps video) | 15 bit/s |
| **High-end security camera** | SNCEB600 Network 720p/30fps HD - Fixed Camera | 1280 x 1024 | 15 bit/s |
| **Extreme camera** | GoPro Hero5 | 4K - WVGA | 100-120 bit/s |
| **Webcam (HD)** | Microsoft LifeCam | 1280x720 (video) | 15 bit/s |
| **Smartphone camera** | Samsung Galaxy S6 | 1920x1080 (video) | 15-60 bit/s |

| Wearable camera | Google Glass Explorer Edition | 2528x1856 1280x 720 (video) | 15 bit/s |
|---|---|---|---|
| **Photodiode sensor** | e.g., SIEMENS photodiode SFH-2030 / Thorlabs PDA100A sensor | - | <4000 bit/s |

*C. Distances*

It is known that the activity of computer LEDs can be detected from more than 30 meters away [27]. In fact, in LED to LED communication, given a line of sight with the transmitter it is possible to detect the LED transmission from even farther away [41]. The quality of the optical LED signal (as received by the camera) is affected by the receiver's location (its distance away, angle, and position), the ambient light, LED wavelength, and other factors. Comprehensive analysis of optical signals is a complex task which goes beyond the scope of this paper. Notably, the environmental conditions can directly influence the effective distance. For example, during the night we were able identify the LED signals from outside the building, at a distance of 20 meters away. In addition, using optical zoom lenses it is possible to extend the range further. Figure 8 shows the signals transmitted from PC-1 as measured in daylight by an optical sensor within the room at distances of three, four, and five meters away. The amplitudes represents the amount of charge converted to voltage with the current amplifier.

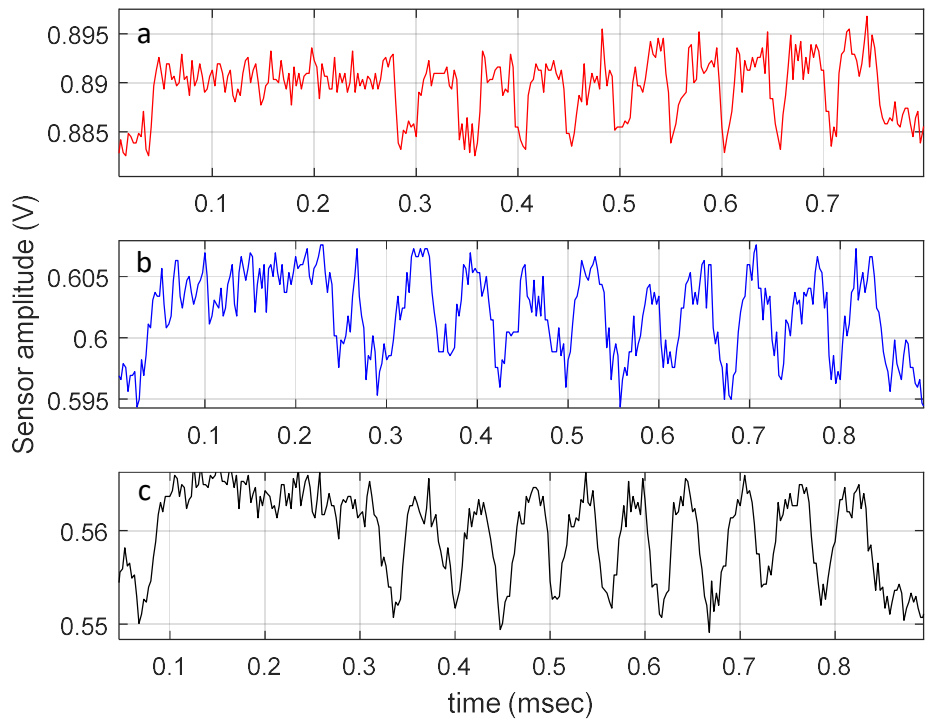

Figure 8. Signals transmitted from PC-1 as measured in daylight by an optical sensor within the room at distances of (a) three, (b) four, and (c) five meters away. The amplitude represents the amount of charge converted to voltage with the current amplifier.

VII. COUNTERMEASURES

Countermeasures for emanation-based data leakage, can be classified into procedural and technological countermeasures. Procedural countermeasures may include organizational practices aimed to restrict the accessibility of sensitive computers by placing them in secured rooms where only authorized staff may access them; typically, all types of cameras are banned from such secured rooms. Some of the NATO standards concerning TEMPEST have been leaked or released [51] [52], and these standards define that certified equipment is classified by zones which refer to the perimeter that needs to be controlled to prevent signal leakage [53]. In these areas, the presence of surveillance cameras may serve as a deterrence measure. However, as mentioned previously in the attack model, the surveillance camera itself may be compromised by a malware [54] [39]. A less sophisticated countermeasure against LED attacks is to disconnect the HDD indication LED or cover it with black tape [27]. Equipment shielding is another commonly recommended by TEMPEST standards. In the case of optical emanation, a special window film that prevents optical eavesdropping may be installed [55]; note that this type of countermeasure doesn't protect against cameras located *within* the building. Technological countermeasures may include the detection of the presence of malware that triggers the HDD activity LED. However, practical implementation of such countermeasures appears to be nontrivial, since the read operation used for the LED control is commonly used by many processes running in the computer. Another possible countermeasure is video monitoring the computer's front panel in order to detect hidden signaling patterns. Again, practical implementation is nontrivial, because the HDD LED routinely blinks frequently due to read and write operations triggered by benign processes. Another interesting solution is to execute a background process that frequently invokes random read and write operations; that way, the signal generated by the malicious process will get mixed up with a random noise, limiting the attack's effectiveness. The list of countermeasures is summarized in Table 8.

Table 8. List of countermeasures

| Type | Countermeasure | Cost |
|---|---|---|
| **Procedural** | Camera banning | Low |
| **Procedural** | LED covering | Low |
| **Procedural** | LED disconnecting | Low |
| **Procedural** | Window shielding | High |
| **Technological** | LED activity monitoring (software) | Low |
| **Technological** | LED activity monitoring (camera) | High |
| **Technological** | Signal jamming (software) | Low |

VIII. CONCLUSION

We present a new method to leak data from air-gapped computers. Our method uses the HDD activity LED which is present in most PC workstations, laptops, and servers today. We show how the malware can indirectly control the status of the LED, turning it on and off for a specified amount of time, by invoking hard drive's 'read' and 'write' operations. Our method is unique in two respects: it is covert and fast. It is covert, because the HDD activity LED routinely blinks frequently; hence, additional blinks caused by the attack may raise no suspicions. In terms of speed, our evaluation shows that the LED is capable of almost 6000 blinks per second, which enables a transmission rate of up to 4000 bit/s. This is 10 times faster than other air-gap covert channels relying on optical emissions. We examine the physical characteristics of HDD LEDs of different colors (red, blue, and white) and tested remote cameras, extreme cameras, security cameras, smartphone cameras, drone cameras, and optical sensors. Our results show that it is feasible to use this optical channel to efficiently leak different types of data (passwords, encryption keys, and files) from an air-gapped computer, via the HDD indication LED.